\documentclass[aps,prd,twocolumn,nofootinbib,longbibliography]{revtex4-1}

\usepackage{times}
\usepackage{amsfonts}
\usepackage{graphicx}
\usepackage{amsmath}
\usepackage{amssymb}
\usepackage{natbib}
  
\begin{document}

 \title{\large Holographic thermodynamics requires a chemical potential for color}
 
\author{   
 Manus R. Visser}

\affiliation{Department of Theoretical Physics, University of Geneva,
24 quai Ernest-Ansermet, 1211 Gen\`{e}ve 4, Switzerland}

\begin{abstract}
The thermodynamic Euler equation for   high-energy states of large-$N$ gauge theories is derived from the dependence of the extensive    quantities on the number of colors $N$. This   Euler equation relates the energy of the state to the   temperature, entropy,   number of degrees of freedom and its     chemical potential, but not to  the       volume or pressure.  
In the context of the  gauge/gravity duality we show that the Euler equation is dual to the generalized Smarr formula for  black holes in the presence of a negative cosmological constant. We also match the fundamental variational equation of   thermodynamics to the     first law of      black hole mechanics, when extended to include variations of the cosmological constant and Newton's constant. 
 \end{abstract}

\maketitle

 \setcounter{secnumdepth}{1}
\section{Introduction}

The thermodynamics of black holes remains one of the most important theoretical advancements in gravitational physics of the past half-century. In semiclassical general relativity  the energy $E$, entropy $S$, and temperature $T$ of a black hole can be identified with its mass, horizon area $A$ and surface gravity~$\kappa$, respectively, \cite{Bekenstein:1972tm,*Bekenstein:1973ur,Hawking:1974sw,Bardeen:1973gs} (setting $k_B = \hbar = c =~1$)
\begin{align} \label{energyentropy}
E  = M   \qquad \quad  S = \frac{A}{4 G} \qquad \quad   T = \frac{\kappa    }{2\pi}  .
\end{align}
  For static, asymptotically flat black holes these thermodynamic  quantities are simply related by the Smarr formula~\cite{Smarr:1972kt} $ M = \frac{d-1}{d-2}\frac{\kappa A}{8 \pi G}$  in   $ d+1 $   spacetime dimensions. The dimension dependent factors    are a peculiar property of the Smarr formula, which typically do not appear in the Euler equation   of thermodynamics relating all extensive and intensive quantities.  Furthermore, for  black holes in the presence of a negative cosmological constant   $\Lambda$  there is an additional term in the generalized Smarr formula   $M = \frac{d-1}{d-2} \frac{\kappa A}{8 \pi G} - \frac{1}{d-2} \frac{\Theta \Lambda}{4 \pi G}$ \cite{Caldarelli:1999xj,Kastor:2009wy}. In an \emph{extended} version of  black hole thermodynamics  the cosmological constant is interpreted as  the    pressure    $P = - \Lambda/ 8\pi G$, and is treated as a thermodynamic state variable in its own right \cite{Kastor:2009wy,Dolan:2010ha,*Dolan:2011xt,Cvetic:2010jb,Kubiznak:2014zwa}. Its conjugate quantity  $\Theta$ in the extended first law of   black holes is   regarded as (minus) the  thermodynamic volume. An important, but  slightly odd     aspect of this interpretation is that the mass of the black hole is identified with the enthalpy   instead of the internal energy of the system (see \cite{Kubiznak:2016qmn} for a recent review).

From a holographic perspective  the dimension dependent factors and the $\Lambda$ term in the generalized Smarr formula remain somewhat elusive. 
 In  holography, or gauge/gravity duality,  the thermodynamics of   black holes in  the  ``bulk''   spacetime is equivalent to the thermodynamics of  large-$N$ strongly coupled gauge theories living on the asymptotic boundary of the bulk spacetime \cite{Maldacena:1997re,Gubser:1998bc,Witten:1998qj}.  In particular,  the  thermodynamic  variables  \eqref{energyentropy} for  black holes correspond with  the   energy $E$, entropy $S$, and temperature $T$ of thermal states in the boundary    theory \cite{Hawking:1982dh,Witten:1998zw}.
The generalized Smarr formula relating these variables in the gravity theory should be dual to the Euler relation for the thermodynamic quantities in the field theory. But the pressure interpretation of the $\Lambda$ term in the Smarr formula does not directly carry over   to the field theory, since the bulk pressure $P$ is not dual to the boundary pressure $p$ and the bulk  thermodynamic volume  is not related to   spatial volume~$V$ of the boundary \cite{Johnson:2014yja}.

In several works~\cite{Kastor:2009wy,Johnson:2014yja,Dolan:2014cja,Kastor:2014dra,Caceres:2015vsa,Karch:2015rpa}    it has been  suggested  that   varying  $\Lambda$  is related   to   varying the number of colors $N$, or   the number of   degrees of freedom $N^2$, in the boundary field theory. For gauge theories   arising from coincident    D-branes, varying $N$ corresponds to varying the number of branes.   
  Further, in   conformal field theories (CFTs)       the number of degrees of freedom is  given by the    central charge~$C$, whose  variation takes us from one CFT to another. 
   In holographic~CFTs dual to Einstein gravity the central charge corresponds to $C 
  \sim L^{d-1}/G$ \cite{Henningson:1998gx,*Henningson:1998ey,Freedman:1999gp,Myers:2010xs,*Myers:2010tj}, 
where  $L$ is  the curvature radius of the bulk geometry, related to the cosmological constant via  $\Lambda = - d(d-1) /  2L^2 $,  and $G$ is     Newton's  constant in $d+1$ dimensions. So varying $C$ in the boundary CFT is dual to varying $\Lambda$ and $G$ in the bulk theory. 
 In addition, it was argued  in  \cite{Karch:2015rpa}   that varying $\Lambda$ in the bulk does not only correspond to varying   $C$ (or $N$), but also to  varying the volume $V$ of the spatial boundary geometry. This is because the bulk curvature radius $L$  is equal to  the boundary curvature radius for a particular boundary metric. We   show, however, that for a different boundary metric varying $\Lambda$  only  corresponds to varying $C$ (and $E$) in the boundary theory, and  not to varying   $V$.  Overall, by building on (and  refining) the holographic  dictionary   in    \cite{Karch:2015rpa}, we propose a precise  boundary description of extended black hole   thermodynamics.

In this paper  
we argue that the    dual field theory description of  black hole thermodynamics   
requires  a chemical potential~$\mu$ for the central charge (see also \cite{Dolan:2014cja,Kastor:2014dra}). 
From  the large-$N$ scaling properties of the field theory we derive the holographic Euler equation
\begin{equation} \label{euler1}
E = T S + \nu^i B_i + \mu  C ,
\end{equation}
and   show that it is holographically dual to the generalized Smarr formula. 
Here   $\nu_i$  are additional chemical potentials for the conserved quantities~$B_i$ (such as charge and angular momentum). As expected, the dimension dependent factors do not feature in~\eqref{euler1}, and the $\Lambda$ term is incorporated in $\mu$. Moreover, $E$ is the standard energy of the field theory  and not the enthalpy. Strikingly though, there is no $pV$ term  in the large-$N$ Euler equation.     We   explain why this is consistent with the fundamental equation of thermodynamics, $dE = T dS - pdV + \nu_i d B_i  + \mu dC $, in which both $V$ and $C$ are varied. Finally,  we   match this  boundary  variational equation   with  the extended  first law of  black holes. 

\section{Thermodynamics of large-$N$ theories} 

 We first derive the   Euler equation from the scaling properties of gauge theories at finite temperature in the  large-$N$ 't~Hooft limit \cite{tHooft:1973alw},   $N \to \infty$ for fixed coupling $\lambda \equiv g^2 N$. Large-$N$    $SU(N)$ gauge theories on compact spaces, with fields in the adjoint representation,  exhibit  a separation between   low-energy states with energy of   $\mathcal O (N^0)$, and  high-energy states  for which the energy scales   as $E \sim N^2$~\cite{Gross:1980he,Witten:1998zw,Sundborg:1999ue}. This is because the low-energy excitations consist of color singlets, whose energy is independent of $N$, and at high energies all the $N^2$ adjoint degrees of freedom contribute on the same footing.  
The low-energy states are in a confined phase and are characterized by a thermal entropy that grows with energy, whereas the  high-energy states   are in a  deconfined phase and behave as a gas of free   particles (at nonzero $\lambda$ there  could   exist an intermediate phase~\cite{Aharony:2003sx}). Other gauge theories at finite temperature   display a  similar (de)confinement phase transition, but the energy in the deconfined phase may scale with a different   power of $N$, e.g., as $N^3$ for an exotic  theory in $d=6$  with $(0,2)$ supersymmetry \cite{Witten:1995zh,*Strominger:1995ac}. 

In     conformal   theories  the central charge $C$ counts the number of   field   degrees of freedom.  For $SU(N)$ gauge theories with conformal symmetry   the central charge scales as $C \sim N^2$ at large $N$,   so high-energy states satisfy $E \sim C$. Since holographic  CFTs are the main examples of large-$N$ theories   we have in mind, we denote the number of degrees of freedom   simply  as $C$ for all  large-$N$ theories.

High-energy states in large-$N$ theories  obey interesting large-$N$ scaling laws and   are dual to black holes in holographic field theories.  By definition the internal energy of these equilibrium states      depends   on     extensive   quantities, such as entropy $S$, volume $V$, and   conserved quantities $B_i$, and on the (intensive) central charge~$C$, i.e., 
$
E =E(S,V,B_i, C).
$
Formally, we can vary the energy with respect to each of these   quantities, while holding the others   fixed. This leads to Gibbs' fundamental   equation of thermodynamics,  
\begin{align} \label{fieldtheoryfirstlaw}
dE &= T dS - p dV + \nu^i d B_i + \mu dC ,
\end{align}
where the temperature $T$, pressure $p$, chemical potentials $\nu^i$, and the chemical potential $\mu$ conjugate to $C$ are defined as
\begin{equation}
\begin{aligned} \label{defintensive}
 T &\equiv \left ( \frac{\partial E}{\partial S} \right)_{V, B_i, C}, \quad p \equiv - \left( \frac{\partial E}{\partial V} \right)_{S, B_i, C},   \\
 \nu^i&\equiv  \left ( \frac{\partial E}{\partial B_i} \right)_{S,V, C} , \quad \mu \equiv \left ( \frac{\partial E}{\partial C} \right)_{S,V, B_i}.
\end{aligned}
\end{equation}
The variation of $C$ in \eqref{fieldtheoryfirstlaw} moves one away from the original field theory content to a theory with a different number of degrees of freedom. On the other hand, for variations which only compare  different thermodynamic states within the same theory, the variable  $C$ is kept fixed.  
Hence, depending on the ensemble, the central charge could be varied or fixed in the fundamental equation of thermodynamics. However,  we observe next that the central charge necessarily has to appear in the large-$N$ Euler relation.

The  entropy and   conserved quantities    scale  with the central charge for   high-energy states,  $S,B_i \sim C$,  reflecting the contribution from all the degrees of freedom.  Thus, the energy function obeys the following scaling relation:  
\begin{equation}
E (\alpha S, V, \alpha B_i, \alpha C)= \alpha  E(S, V, B_i , C),
\end{equation}
with  $\alpha$ being  a dimensionless scaling parameter. This means that  in the deconfined phase of large-$N$ theories on compact spaces the energy is not an extensive function. 
 Differentiating with respect to $\alpha$ and putting $\alpha=1$ leads to the Euler equation
\begin{equation} \label{holographiceuler}
E = T S + \nu^i B_i + \mu C.
\end{equation}
Notice that pressure and volume do not appear in this    Euler equation, since the volume does not generically scale with~$C$. It does scale with $C$ in the infinite-volume limit of   CFTs, i.e., $pV\! =\!- \mu C $ as $V \to \infty$ (see Appendix~\ref{appA}). In that limit the energy becomes an extensive function, satisfying $E(\alpha S,\alpha V, \alpha B_i)= \alpha E(S,V, B_i)$.  By varying the Euler  relation \eqref{holographiceuler} and employing  the fundamental variational equation~\eqref{fieldtheoryfirstlaw}, we find a slightly unusual Gibbs-Duhem equation
\begin{equation}
0 = S dT  + p dV + B_i d \nu^i + C d \mu.
\end{equation}
The variation of   volume (instead of   pressure) features in this equation, since the Euler relation does not involve a $pV$ term.

 Furthermore, in the grand canonical ensemble the thermodynamic potential or free energy is defined as
\begin{equation} \label{grandcanfree}
W \equiv E - TS - \nu^i B_i  = \mu C.
\end{equation}
It follows    from  the fundamental equation \eqref{fieldtheoryfirstlaw} that the grand canonical  free energy is stationary at fixed $(T, V,\nu^i,C)$.  The proportionality of free energy with $C$  (or $N^2$) is a signature of deconfinement; in contrast, the free energy of the confined phase is of   order one \cite{Thorn:1980iv}.  In fact, the relation $W \sim C$ can be viewed as the definition of the dimensionless central charge $C$ in this paper, which could hence be called the ``thermal free energy charge.'' This charge is generically not identical to other definitions of the central charge, such as anomaly coefficients or the
coefficient of the two-point function of the stress-energy tensor, except in the special case of $d=2$ and in the large-$N$ limit of $SU(N)$ conformal gauge theories (where all central charges scale as $N^2$).
 
The   Euler equation \eqref{holographiceuler}, or equivalently $W =\mu C$,  only holds      in a regime where $1/C$ corrections can be neglected. For generic CFTs on compact spaces this is   the case in the high-temperature or large-volume regime $T R \gg 1$, where $R$ is the curvature radius, since the free energy satisfies $W \sim (TR)^{d-1}$ in that regime and the central charge $C$ is defined as the dimensionless proportionality coefficient.    On the other hand, for holographic    and $2d$~sparse CFTs the free energy already scales with   the  central charge       at low temperatures $T R \sim \mathcal O(1)$ (i.e., if $E R \sim C$ with $C \gg 1$). Further,   the Euler equation is satisfied    for any value of~$\lambda$,   at weak and strong  coupling, and for any large-$N$ field theory, including  conformal and   confining theories, and theories with unusual scaling behavior  like  Lifshitz     theories. Each of these theories, though,  satisfies a different equation of state, which is not encoded in the  large-$N$ Euler equation~\cite{Karch:2015rpa}. For instance,  the equation of state for conformal theories is $E = (d-1) pV  $, and   for  Lifshitz scale  invariant theories   with   dynamical scaling exponent $z$  it  is given by $ z E = (d-1  ) pV$ (see Appendix~\ref{appA}). The fact that the  Euler relation   applies to both conformal and Lifshitz theories, means that it   not only holds for relativistic,  but also for nonrelativistic theories.

\section{Holographic black hole thermodynamics} The large-$N$ Euler equation  applies in particular to strongly coupled large-$N$ CFTs with a semiclassical, gravitational dual  description. We now investigate the gravity dual of the Euler equation.  

The best-established example of holography, the  anti-de Sitter/conformal field theory (AdS/CFT)  correspondence, states that the   partition function of the CFT and of  the gravitational theory in asymptotically AdS spacetime    are equal 
$
Z_{\text{CFT}}   = Z_{\text{AdS}}  
$~\cite{Gubser:1998bc,Witten:1998qj}. 
For   field theories at finite temperature  the thermal partition function is related to the free energy via $W =- T \ln Z_{\text{CFT}}$. On the other hand, the gravitational partition function is given by the Euclidean path integral, which  in the saddle-point approximation is computed by  the on-shell Euclidean action, $I_{\text{E}} =  - \ln Z_{\text{AdS}}$~\cite{Gibbons:1976ue}.   Since thermal states in the CFT are dual to black holes in AdS, the on-shell action should be evaluated on the classical black hole saddle. Here, we consider   rotating, charged black hole solutions \cite{Caldarelli:1999xj} to the Einstein-Maxwell action with a negative cosmological constant, i.e. $I_\text{E} = - \frac{1}{16\pi G} \int d^{d+1} x \sqrt{ g}(R - 2 \Lambda - F^2)$. In the grand canonical ensemble (at fixed $T$ and~$\Phi$) the free energy of the holographic field theory  corresponds  to   \cite{Chamblin:1999tk,Gibbons:2004ai,Papadimitriou:2005ii}
\begin{equation} \label{adscftdic}
W = T I_{\text{E}}= M - \frac{\kappa A}{8\pi G} - \Omega J - \Phi Q.
\end{equation}
The final equality follows from evaluating the action --- including the Gibbons-Hawking boundary term \cite{Gibbons:1976ue} and a background subtraction term --- on the black hole solution with angular momentum $J$ and electric charge~$Q$. The corresponding   chemical potentials are the angular velocity $\Omega$ and the electric potential $\Phi$ of the   horizon.   We note   it is straightforward to generalize this equation to black holes with multiple electric charges and angular momenta \cite{Myers:1986un,Hawking:1998kw,Gibbons:2004js,*Gibbons:2004uw}.   
 
The thermodynamic Euler equation for these black holes follows from \eqref{adscftdic} by inserting the   holographic dictionary \eqref{energyentropy} for energy, entropy, and temperature, and the dictionary for the charge $\tilde Q = Q L $ and potential $\tilde \Phi = \Phi/ L$ \cite{Chamblin:1999tk,Karch:2015rpa}, and using the  relation~\eqref{grandcanfree} between   free energy and    chemical potential, 
  \begin{equation}  \label{euler}
 	E = T S + \Omega J  + \tilde \Phi \tilde Q + \mu C .
\end{equation}
The thermodynamic variables in this equation are well-known black hole parameters, except for the chemical potential and central charge. What is their gravitational dual description?   Essentially, $\mu C$ is the on-shell Euclidean action (times $T$). In addition, a different expression for the chemical potential is obtained from the generalized Smarr formula for AdS black holes, which relates all the black hole parameters and is thus a gravitational  reorganization of the Euler relation \cite{Kastor:2009wy,Caldarelli:1999xj,Dolan:2010ha,*Dolan:2011xt,Cvetic:2010jb,Gibbons:2004ai,Barnich:2004uw},  
\begin{equation} \label{smarrstatic}
  M  = \frac{d-1}{d-2} \left (  \frac{\kappa  A}{8 \pi G} +  \Omega  J \right)  +   \Phi Q   -  \frac{1}{d-2}\frac{  \Theta\Lambda}{ 4\pi G } .
\end{equation}
The $\Lambda$   term  is absent for asymptotically flat black holes, but is   necessary for the consistency of the Smarr formula of asymptotically AdS black holes.  The quantity  $\Theta$ can be defined as $\int_{\Sigma_{\text{bh}}} |\xi| dV - \int_{\Sigma_{\text{AdS}}} |\xi| dV  $ \cite{Jacobson:2018ahi}, where a subtraction with respect to the pure AdS background is implemented to cancel the divergence at infinity. In this definition the domain of integration $\Sigma_{\text{BH}}$ extends from the horizon to infinity, while    $\Sigma_{\text{AdS}}$ in the pure AdS integral extends across the entire spacetime. Further, $\xi$ is the timelike Killing field $\xi = \partial_t + \Omega  \partial_{\phi}$, which (in the black hole geometry) generates the event horizon, and $|\xi| = \sqrt{-\xi \cdot \xi}$ is its norm.  In the literature  \cite{Dolan:2010ha,*Dolan:2011xt,Cvetic:2010jb} (minus) $\Theta$ is often   called the ``thermodynamic volume,'' since it is the conjugate quantity to $\Lambda$ (the bulk pressure) in the first law of black hole mechanics, see Eq.~\eqref{staticfirstlaw}. For our   purposes, however, a  geometric name is probably more suitable, such as (background subtracted) ``Killing volume,'' because we are interested in the field theory thermodynamics rather than the bulk thermodynamics. 

Comparing the Euler equation and the Smarr formula we see that the   chemical potential (times central charge)   corresponds to   three combinations of the black hole parameters
 \begin{align} \label{chemicalstatic}
\mu C &= M - \frac{\kappa A}{8\pi G} - \Omega J - \Phi Q =\frac{1 }{d-1} \!\!\left ( M - \Phi Q -   \frac{  \Theta \Lambda  }{ 4\pi G} \right) \nonumber \\
&=\frac{1}{d-2} \left (\frac{\kappa A}{8 \pi G} +  \Omega   J  - \frac{\Theta \Lambda}{4 \pi G}\right).
\end{align}
Note that the dimension dependent factors in the Smarr formula are absorbed in the chemical potential. 
The expression above for the chemical potential should also follow from its definition   in \eqref{defintensive},
$
	\mu \equiv \left ( \frac{\partial E}{\partial C} \right)_{S,V,J, \tilde Q}  . 
$ 
 We check this explicitly  by rewriting the extended first law of AdS black hole mechanics as a thermodynamic variational identity, where $\mu$  plays the role of   the   conjugate quantity to the central charge variation $dC$. For CFTs dual to Einstein gravity the  holographic dictionary for the central charge depends on both the cosmological constant $\Lambda$ and Newton's constant $G$. In order to keep track of the central charge variation,   we vary both coupling constants as ``bookkeeping devices'' in the   bulk first law \cite{Kastor:2014dra,Karch:2015rpa}. 
 
 The mass of rotating, charged AdS black holes  can be regarded as the function  $  M (A, J, Q, \Lambda, G)$.    From a   scaling argument~\cite{Smarr:1972kt,Kastor:2009wy}   and from  the   generalized Smarr  formula \eqref{smarrstatic} it follows that the extended first law   for these  black holes is 
 \begin{equation}
 \begin{aligned} \label{staticfirstlaw}
 	d M &= \frac{\kappa}{8\pi G} dA + \Omega  dJ  + \Phi dQ   + \frac{\Theta}{8\pi G} d \Lambda \\
 	&-  \left ( M - \Omega  J - \Phi Q  \right) \frac{dG}{G}.
 \end{aligned}
 \end{equation}
Usually, in extended black hole thermodynamics   only the variation of $\Lambda$ is taken into account in the first law, but the variation of Newton's constant can   be easily included    by noting that $M, J, Q \sim G^{-1}$ \cite{Kastor:2010gq,Sarkar:2020yjs}. 
 Remarkably, the   $\Lambda$ and $G$ variations in \eqref{staticfirstlaw} cannot be combined into one single term   proportional to $d(\Lambda/ G)$, because there is     a  term remaining involving the variation of $G$. This seems to imply that the standard  interpretation of the extended first law in terms of bulk pressure $P=-\Lambda/8\pi G$ is inconsistent, if Newton's constant is allowed to vary. On the other hand, we can find a consistent boundary   interpretation by expressing the right-hand side of the first law in terms  of     variations of the  entropy  $S= A/4 G$, electric charge $\tilde Q = QL$, central charge $C \sim L^{d-1}/ G$, and   spatial volume  $V \sim L^{d-1}$ of the holographic field theory. To this end we   rewrite the extended first law   above     as
  \begin{align} \label{firstlawL}
 	d M &= \frac{\kappa}{2\pi} d \left (\frac{A}{4G} \right) + \Omega  dJ   + \frac{\Phi}{L}  d (Q L) - \frac{M}{d-1} \frac{dL^{d-1}}{L^{d-1}} \nonumber \\
 	&+  \left (  M   -   \frac{\kappa  A}{8 \pi G}   -  \Omega  J   - \frac{\Phi}{L} QL  \right)  \frac{d (L^{d-1}/G)}{L^{d-1}/G}. 
 	 \end{align}
Here we used again the Smarr relation   and $d \Lambda / \Lambda = -2 d L / L$,    
and observed that   $(d-1) dL / L - d G / G$ is equal to the fraction in the final term. It is crucial that the $L$ and $G$ variations appear in this combination, otherwise the  holographically dual first law would not involve a variation of the central charge.  Moreover, by allowing for variations of $G$ we    can clearly  distinguish the variation of the spatial volume $V$ from that of the central charge $C$  \cite{Karch:2015rpa}. 
Consequently, from the holographic dictionary  
we deduce that the extended first law for AdS black holes is dual to the fundamental   equation in thermodynamics, 
\begin{equation} \label{cftfirstlaw}
 	d E = T dS + \Omega  d J  + \tilde \Phi d \tilde Q - p dV  + \mu dC .
\end{equation}
By comparing \eqref{firstlawL} and \eqref{cftfirstlaw} we see that  the pressure~$p$ satisfies the CFT equation of state, $E = (d-1) p V$, and the chemical   potential $\mu$ fulfils the Euler equation \eqref{euler}. This shows that our dictionary \eqref{chemicalstatic} for the chemical potential   is consistent, since the same expression follows from equating bulk and boundary free energy and from matching the ``first laws.'' Note that we only used the scaling properties $V \sim L^{d-1}$ and $C \sim L^{d-1}/G$ to arrive  at \eqref{cftfirstlaw}, but did not need their proportionality constants, because   the fractions $dV/V$ and $dC/C$  appear   in the first law and hence the  proportionality constants  drop out. 

As an aside, we mention that the precise match between the  first laws  can be generalized to charged Lifshitz black holes~\cite{Tarrio:2011de}  by replacing the   holographic dictionary with: $ E = M L^{1-z} , \, T =  (\kappa/2\pi) L^{1-z}$ and $\tilde \Phi = \Phi / L^z$ (see   Appendices~\ref{appA}~and~\ref{appDnew}).  The extended first law \eqref{staticfirstlaw} with $J=0$   still holds for Lifshitz black holes \cite{Brenna:2015pqa}, and is dual to the   fundamental equation \eqref{cftfirstlaw}, if  $p$   satisfies the Lifshitz   equation of state   $z E = (d-1)pV$ and $\mu$ satisfies the Euler relation \cite{Vissernew}.

\section{Comparison with previous literature} In Refs. \cite{Dolan:2014cja,Zhang:2014uoa,*Zhang:2015ova} the chemical potential associated to the central charge (or $N^2$) was defined as $\mu \equiv \left ( \frac{\partial E}{\partial C} \right)_{S}$ for AdS-Schwarzschild black holes. Compared to our Eq.  \eqref{defintensive} the fixed volume requirement is lacking in this definition.  From the  extended first law it follows that this definition of the chemical potential is   proportional to the Killing volume~$\Theta$, which is inconsistent with  the Euler equation and  our expressions for $\mu$ in  \eqref{chemicalstatic}.  
 This definition is especially problematic since these references take the boundary curvature radius  $R$   to be equal to the bulk curvature radius~$L$, which implies that the spatial volume and   central charge are both proportional to $L^{d-1}$. Thus, in \cite{Dolan:2014cja,Zhang:2014uoa,*Zhang:2015ova}  they mix up the bulk duals to spatial volume and central charge.  

In \cite{Karch:2015rpa} this issue was resolved by allowing for variations of~$G$, in addition to  $\Lambda$, so that the central charge variation can be distinguished from the volume variation.  Our matching  above  between the  first laws   is based on this approach, but is still novel since \cite{Karch:2015rpa}    focused on finding the boundary dual to the Smarr relation and not to  the first law.    In \cite{Karch:2015rpa} 
the free energy relation  $W \sim N^2$ at large-$N$, which is equivalent to our Euler equation \eqref{grandcanfree}, was     identified as the holographic origin of the   Smarr formula. In addition, the pressure and its equation of state played an important role in their holographic derivation of the Smarr formula. However, their proof only holds for a particular   choice of CFT  metric, $ds^2=-dt^2 + L^2 d \Omega_{k, d-1}^2$, where  $L$   is   the AdS radius (see Appendix \ref{appC}). The derivation can be extended to a more general CFT metric with $R \neq L$ and, remarkably, it does not depend on the boundary pressure in this case. Rescaling the CFT metric above with  the Weyl factor $\lambda = R/L$  changes the CFT time into $R t / L$ and the boundary curvature radius into~$R$, so that  the spatial volume becomes $V \sim R^{d-1}$.  The added benefit of this more general boundary metric is   that $V$ is   clearly  distinct from $C$ (and $ \tilde Q$).  The  (refined) holographic  dictionary for this metric   is \cite{Savonije:2001nd}
  \begin{equation} \label{rescaleddict}
E =  M \frac{ L}{R}, \quad T = \frac{\kappa}{2\pi} \frac{L}{R}, \quad  \tilde \Omega = \Omega \frac{L} {R} ,\quad \tilde \Phi = \frac{\Phi}{R}.
\end{equation}
Importantly,  with this dictionary the bulk and boundary variational equations  \eqref{staticfirstlaw}  and     \eqref{cftfirstlaw}  still agree    and the chemical potential  again satisfies the Euler relation.   We can even keep $G$ fixed    and only vary $\Lambda$ in the bulk, since $dV \sim  dR^{d-1}$ and $d C \sim dL^{d-1} \big|_G$ are   obviously   distinguishable  for $R \neq L$.

  Now, we derive the bulk Smarr formula purely   from the boundary Euler equation and the holographic dictionary  for $R \neq L$. The $\Lambda$~term in the Smarr formula  can be expressed as
\begin{equation}
- \frac{\Theta \Lambda}{4 \pi G} = L \left ( \frac{\partial M }{\partial L} \right)_{\!\!A, J, Q, G} \!\!= R \left (\frac{\partial E}{\partial L} \right)_{\!\!A, J, Q, G}\!\! - E \frac{R}{L} .
 \end{equation}
Note that the bulk quantities $A, J, Q$ and $G$ are kept fixed in the partial derivative. 
The boundary energy depends implicitly on them  as:  $E = E(S (A,G),  J, \tilde Q (L, Q), V (R), C(L, G))$. This implies (see Appendix \ref{appDnew} for more details)
\begin{equation}
- \frac{\Theta \Lambda}{4 \pi G} =  \frac{R}{L} \left (  \tilde \Phi \tilde Q + (d-1) \mu C - E \right),
\end{equation}
where we used the definitions of  $\tilde \Phi$ and $\mu$ from \eqref{defintensive}. Finally, inserting the Euler equation and the holographic dictionary \eqref{rescaleddict}
  precisely yields the Smarr formula \eqref{smarrstatic}.  Note that this derivation hinges on the     chemical potential   and not on the pressure.

\section{Discussion}   In gauge/gravity duality, black holes in the bulk  correspond to thermal states   in the boundary theory.  
 We proposed a new dictionary  between the bulk and boundary thermodynamics  by introducing a chemical potential for the number of colors in the gauge theory. The chemical potential is crucial for the  correspondence between the Euler equation for large-$N$ theories and  the Smarr formula relating the black hole parameters.  
Since the Euler relation determines the energy  as a function of other   variables, it contains the essential thermodynamic information about the field theory.  

 Our   field theory interpretation of  the  extended  thermodynamics of black holes stands in contrast to the common gravitational interpretation   in terms of bulk pressure and volume.  One notable difference is that the black hole mass is equivalent to the internal energy of the field theory, whereas   in  \cite{Kastor:2009wy} it is identified   with the enthalpy of the gravitational system. Moreover, we found that the extended first law of black holes cannot be solely written in terms of the variation of bulk pressure, $P=-\Lambda /8\pi G$, if both $\Lambda$ and $G$ are allowed to vary, but can be consistently interpreted as a field theory first law. Thus, as the thermal field theory  has a natural thermodynamic description, the boundary interpretation     seems unavoidable.

As for future work, we expect that the  dictionary for the chemical potential can be generalized to a  multitude of black holes in the presence of a cosmological constant, such as black holes in    higher-curvature gravity,     hyperscaling violating solutions, black rings      and  de Sitter black holes. On the field theory side, an  interesting problem is to extend the Euler equation   beyond the large-$N$ limit, by including $1/N$ corrections \cite{Karch:2015rpa,Sinamuli:2017rhp}.

 \section*{Acknowledgments} I would like to thank   Jan de Boer, Pablo Bueno, Matthijs Hogervorst, Arunabha Saha and Watse Sybesma     for useful  discussions, and Andreas Karch, Juan Pedraza, Andrew Svesko and an anonymous referee   for detailed comments on a   draft. This work was supported   by the  Republic and canton of Geneva and   the Swiss National Science Foundation, through Project Grants No. 200020-182513 and  No. 51NF40-141869  The Mathematics of Physics (SwissMAP).

  \appendix

\section{Euler equation for   two-dimensional CFTs} 
\label{app2dcft}

Examples of  large-$N$  field  theories are $2d$   modular invariant CFTs with   large central charge $c$. The microcanonical entropy for these theories is given by the Cardy formula (setting $c_L = c_R = c$)~\cite{Cardy:1986ie}
\begin{equation} \label{cardy}
S (E_L, E_R, c) = 2 \pi \sqrt{\frac{c}{6} E_L} + 2 \pi \sqrt{\frac{c}{6}E_R} ,
\end{equation}
with $E_{L,R}$     the left- and right-moving energies. On a circle of length $V=2\pi R$, the total energy and angular momentum  are, respectively, $E  = (E_L + E_R )/R$ and $J = E_L - E_R$. The Cardy formula holds for CFTs with a sparse light spectrum  in the   regime $C \to \infty$ with   $E R  \ge C$  \cite{Hartman:2014oaa}, where we normalized the central charge (conjugate to~$\mu$)   as $C=c/12$.  If we view the entropy \eqref{cardy} as the function   $S= S(E, V,J, C)$,   then  
the fundamental variational equation  of thermodynamics with $\nu^i dB_i = \Omega dJ$, follows by taking partial derivatives of the entropy function. Consequently, the products of  thermodynamic   quantities  are
\begin{equation}
\begin{aligned} \label{2dcftdict}
	T S &= \frac{4}{R}  \sqrt{E_L E_R} ,\qquad   \, \qquad  p V = E,  \\
	 \Omega J&= E - \frac{2}{R} \sqrt{E_L E_R},  \qquad \mu C  = -\frac{2}{ R} \sqrt{E_L E_R } ,
\end{aligned}
\end{equation}
where $\Omega$ is the angular potential. 
They satisfy the    relation
\begin{equation} \label{eq:euler2d}
	E = T S + \Omega J + \mu C .
\end{equation}
Hence, the large-$N$ Euler equation indeed holds for $2d$ CFTs. In fact,   the Euler relation  splits up into two separate equations,   $E = \Omega J - \mu C $ and $TS =- 2  \mu C  $. 

In   AdS$_3$ gravity the   Smarr formula  for the outer   horizon of a BTZ black hole is given by  $0 =  TS + \Omega J -   \Theta \Lambda / 4 \pi G$ \cite{Frassino:2015oca}. Comparing this to the Euler equation \eqref{eq:euler2d} we find that the chemical potential must correspond to $\mu C = E - \Theta \Lambda/ 4\pi G$. Using the  holographic dictionary for the central charge $c=3L/2G$~\cite{Brown:1986nw,Strominger:1997eq}, it can be shown that the chemical potential   is dual to 
$
	\mu \!   
	= \!- ( r_+^2 - r_-^2)/ ( L^2 R) , 
$ 
where $r_\pm$ are the outer and inner horizon radii of the rotating  BTZ black hole. 
Notably, $\mu$ vanishes for extremal   black holes, if $r_+ = r_-$ or $ E R = |J|$, which correspond  to CFT states with $E_L=0$ or $E_R=0$.

\section{The extended first law of entanglement}

In this appendix we compare our chemical potential   for     AdS black holes to the chemical potential in the extended first law for   entanglement entropy of ball-shaped regions in the CFT vacuum \cite{Blanco:2013joa,*Wong:2013gua,Kastor:2014dra}. This CFT first law takes the form 
\begin{equation}
	d\bar E = \bar T dS_{\text{ent}} + \bar \mu dC,
\end{equation}
where $\bar E$ denotes the modular Hamiltonian expectation value, $S_{\text{ent}}$ is the vacuum entanglement entropy of the ball-shaped region and $C$ is the universal coefficient of the entanglement entropy (commonly denoted as $a^*_d$) \cite{Myers:2010xs,*Myers:2010tj,Caceres:2016xjz,Rosso:2020zkk}. The CFT     first law is    dual to the first law of      static hyperbolic AdS black holes  which are isometric to pure AdS space \cite{Casini:2011kv,Faulkner:2013ica,Emparan:1999gf},  a special case of the black holes considered in the main text, with $J=Q=0$. The  boundary first law     follows from reformulating our fundamental  variational equation  in terms of   dimensionless   quantities $\bar E = M L, \bar T = \kappa L / 2\pi$, and $ \bar \mu = \mu L  $. The volume variation drops out of the first law, since it is a dimensionful quantity.  
In the vacuum $\bar E=0$, hence  the chemical potential  reduces to $\bar \mu = - \bar T S_{\text{ent}} / C$, which agrees with the results in  \cite{Kastor:2014dra}   (where the temperature was normalized as $\bar T=1$).



 \section{Euler equation in flat space}
\label{appA}

In flat spacetime,    static   equilibrium  states satisfy   the standard    thermodynamic Euler equation, 
\begin{equation}
E = T S + \nu^i B_i - p V ,
\end{equation}
which is often formulated instead  in terms of densities since $V$ is infinite. Note that the energy is purely extensive in this formula, since it satisfies $E (\alpha S,  \alpha  V, \alpha B_i) = \alpha E (S,  V, B_i)$. This Euler relation  applies in particular to conformal and Lifshitz theories on the plane (see e.g.~\cite{Natsuume:2014sfa,Taylor:2015glc}). It is not   immediately clear why this equation is consistent with the large-$N$ Euler equation,  therefore  in this appendix we   explain the relation between the two for   Lifshitz  scale invariant  theories.   

Anisotropic scaling symmetry $\left  \{ t,  x^i \right \} \to \left  \{\zeta^z t, \zeta x^i \right \}$ with      dynamical scaling exponent $z$ implies that   the   product $T R^z $ is  Lifshitz scale invariant, where $R$ is the curvature radius of the compact space, such as a sphere. Therefore, for Lifshitz      theories with positive~$z$ the infinite-volume limit $R \to \infty$ is   effectively the same as  $T \to \infty$, so on the plane these theories are essentially always in the high-temperature deconfining phase.  
  In this limit, the energy scales as $E \sim T^{ \frac{d-1   + z }{z}  }$ and entropy and conserved quantities  as $S, B_i \sim T^{ \frac{d-1 }{z}  }$, so  the    scaling relation is  $E(\alpha^{ \frac{d-1}{z}  } S, V, \alpha^{\frac{ d-1}{z}  } B_i, C) = \alpha^{\frac{d-1 + z }{z} }E (S, V, B_i , C)$. This    imposes the  condition   $(d-1  +z  )E = (d-1  )(TS + \nu^i B_i)$, which in combination with  the large-$N$  Euler equation yields  $z E = - (d-1   )\mu C$. We can now compare this to the Lifshitz equation of state $z E =  (d-1   ) p V$, which is a consequence of the anisotropic   scaling relation  $E (S, \alpha^{d-1   } V, B_i, C) = \alpha^{-z  } E(S, V, B_i,C)$. As a result, we find    $\mu C = - pV$ as $V \to \infty$,  turning  the large-$N$ Euler equation into the  standard one. The same argument works for conformal theories (by setting $z=1$),  hyperscaling violating theories   and possibly        other large-$N$ theories.  
  Notably,   the standard   Euler equation only applies  in the infinite-volume limit of   large-$N$ theories. The large-$N$ Euler relation, on the other hand, also holds  at finite temperature on compact spaces for holographic field theories and   $2d$ sparse CFTs  (but not for generic CFTs).

     \section{Holographic derivation  of the Smarr formula for Lifshitz black holes}
     \label{appDnew}

In this appendix we derive the Smarr formula  for  charged Lifshitz black holes \cite{Tarrio:2011de}, with curvature radius $L$ and  scaling exponent $z$,   from the holographic Euler equation and the dictionary for the thermodynamic quantities involved.  We put the dual Lifshitz field theory on  a spatial geometry of  curvature radius~$R$.   Our derivation generalizes  section 2.3 of  \cite{Karch:2015rpa}  to $R \neq L$ and $z \neq 1$. Our aim is to prove that  even for Lifshitz black holes the boundary pressure and   its equation of state are  not necessary input to deduce the Smarr formula (although they are in the special case $R = L$ considered in \cite{Karch:2015rpa}). 

The holographic dictionary for Lifshitz black holes reads
\begin{align} \label{dictionarylifshitz}
E = M \frac{L}{R^z}, \quad T = \frac{\kappa}{2\pi} \frac{L}{R^z}, \quad \tilde \Phi = \frac{\Phi}{R^z}, \quad \tilde Q = Q L .
\end{align}
Note that the factors of $R$ and $L$ are chosen such that   the products $E R^z  $, $T R^z$ and $\tilde \Phi R^z$ are Lifshitz scale invariant (see Appendix \ref{appA}). 
First,  we  express the $\Lambda$ term in the   Smarr formula in terms of  the  boundary  energy  $E$
\begin{equation} \label{lifshitzkillingvolume}
- \frac{\Theta \Lambda}{4 \pi G} = L  \left ( \frac{\partial M }{\partial L}  \right)_{\!\!A, Q, G} \!\!= R^z \left ( \frac{\partial E}{\partial L } \right)_{\!\!A, Q, G} - E \frac{R^z}{L} .
\end{equation}
The strategy is to show that the right-hand side satisfies the Smarr formula. Note that the bulk quantities $A, Q$ and $G$ are fixed in the partial derivative with respect to   $L$.  
The boundary energy depends on  these  bulk quantities as follows 
\begin{equation} \label{energyfunction}
E = E(S(A,G),  \tilde Q (L,Q), V(R), C (L, G)).
\end{equation}
Note that $J=0$. The partial derivative is hence   given by
\begin{align} \label{intermediatestep}
\left (  \frac{\partial E}{\partial L } \right)_{\!\!A, Q, G} \!\! &= \left(  \frac{\partial E}{\partial \tilde Q} \right)_{\!\!S, V, C}\!\! \left (  \frac{\partial \tilde Q}{\partial L} \right)_{\!\!Q} \!\! +\left ( \frac{\partial E}{\partial C} \right)_{\!\! S, V, \tilde Q}   \left ( \frac{\partial C}{\partial L} \right)_{\!\!G} \nonumber \\
&= \frac{1}{L} \! \left (  \tilde \Phi  \tilde Q +(d-1) \mu C \right).
\end{align}
In the second line  we used $\tilde Q = QL$ and $C \sim L^{d-1}/G$, and  we recognized the definitions of  the  electric potential $\tilde \Phi$ and   chemical potential $\mu$ (see  Eq. \eqref{defintensive} in the main text). 
Thus, we find
\begin{align} \label{smarrderivation}
- \frac{\Theta \Lambda}{4 \pi G}  &= \frac{R^z}{L} \! \left (    \tilde \Phi  \tilde Q +(d-1) \mu  C  -E   \right)    \\
&= \frac{R^z}{L} \! \left (    (d-2) E - (d-1) TS - (d-2) \tilde \Phi  \tilde Q     \right) .  \nonumber
\end{align}
Finally, by inserting the holographic dictionary \eqref{dictionarylifshitz} we   recover the   Smarr formula. Note that the Smarr formula for Lifshitz black holes does not involve $z$ and is hence the same as for black holes in Einstein gravity  \cite{Brenna:2015pqa}.  Crucially, the holographic Euler equation was employed in the second line of \eqref{smarrderivation} and is therefore dual to the Smarr formula, as pointed out in \cite{Karch:2015rpa}.  We emphasize that the boundary pressure does not play a role in this derivation, whereas the chemical potential $\mu$ does.  For $R=L$ the pressure does feature in the derivation, since in that case the boundary volume   depends on the bulk radius, i.e.  $V (L)$, which yields an extra term $-(d-1) pV / L$ in \eqref{intermediatestep}. But ultimately  the result in \eqref{smarrderivation} remains the same, since this pressure term cancels, due to the Lifshitz equation of state,  against a new term $   z E L^{z-1} $ on the right side of   \eqref{lifshitzkillingvolume}.

\section{The  renormalized holographic Euler equation}
\label{appC}

In the main text the   energy was defined with respect to the ground state,  so   the vacuum energy was effectively set to zero. However, CFTs on a curved background exhibit the Casimir effect, which implies that   the ground state could have   nonvanishing energy. In AdS/CFT the ground-state energy can be computed   with the method of holographic renormalization,  by regularizing the gravitational action with local counterterms at the boundary \cite{Henningson:1998gx,Balasubramanian:1999re}. In this appendix we derive the renormalized holographic Euler equation for static vacuum AdS black holes, and find that the ground-state energy contributes a constant term to the chemical potential.

We consider   static, vacuum asymptotically AdS black holes with hyperbolic, planar and spherical horizons   \cite{Birmingham:1998nr}
\begin{equation}
	ds^2 = -f_k (r) dt^2 + \frac{dr^2}{f_k(r)} + r^2 d \Omega_{k,d-1}^2,
\end{equation}
where 
\begin{equation}
	f_k (r) = k + \frac{r^2}{L^2} - \frac{16 \pi G M }{(d-1) \Omega_{k,d-1}r^{d-1}}  .
\end{equation}
 For $k=1$  the unit metric $d \Omega_{k,d-1}^2 \!$ is the metric on a unit $S^{d-1}\!$ sphere, for $k=0$ it is the dimensionless metric $\frac{1}{L^2} \!\sum_{i=1}^{d-1} dx_i^2  $    on the plane $\mathbb R^{d-1}$, and for $k=-1$ the unit metric on hyperbolic space $H^{d-1}$ is $  d u^2 + \sinh^2 u d \Omega_{k=1,d-2}^2$.   
 The mass  parameter $M$   is related to the horizon radius $r_+$ via  
 \begin{equation}
 	M = \frac{(d-1) \Omega_{k,d-1} r_+^{d-2}}{16 \pi G } \left ( \frac{r_+^2}{L^2} + k    \right).
 \end{equation}
   According to the Gubser-Klebanov-Polyakov-Witten prescription in AdS/CFT \cite{Gubser:1998bc,Witten:1998qj},  the CFT metric is   identified with the boundary metric  of the dual asymptotically AdS spacetime up to a Weyl rescaling,  i.e., $g_{\text{CFT}}= \lim_{r \to \infty}\lambda^2 (x)g_{\text{AdS}}  $  where $\lambda(x)$ is a Weyl scale factor.  As   $r\to \infty$ the boundary metric  
   approaches 
  \begin{equation}\label{asymptads}
  	ds^2 = \frac{r^2}{L^2} dt^2 + \frac{L^2}{r^2} dr^2  + r^2 d\Omega_{k,d-1}^2.
  \end{equation}
A common choice of   Weyl factor  is  $\lambda =L /r$, so that   the CFT metric becomes $- dt^2 + L^2 d \Omega_{k,d-1}^2$. The   boundary curvature radius   is then equal to the AdS radius and  the    volume is $V = \Omega_{k,d-1} L^{d-1}/(d-1)$.  Moreover, the CFT time is the same as the  global AdS time $t$, which implies that the CFT energy $E$ can be identified (up to a constant) with the ADM mass~$M$, the conserved charge associated to time $t$ translations.

  The temperature, entropy and energy of the  black holes are 
 \begin{align} \label{appenergy}
 	T &=\frac{d \, r_+^2 + k (d-2)L^2}{4 \pi L^2 r_+}, \qquad  
 	S= \frac{\Omega_{k,d-1}r_+^{d-1}}{4 G}, \\
 	E_{\text{ren}}  &= \frac{(d-1) \Omega_{k,d-1} L^{d-2}}{16 \pi G } \left ( \frac{r_+^d}{L^d} + k \frac{r_+^{d-2}}{L^{d-2}} + \frac{2 \epsilon_k^0}{d-1} \right). \nonumber 
 	\end{align}
 	The energy was derived from the renormalized boundary stress-energy tensor in \cite{Balasubramanian:1999re} and from the on-shell Euclidean gravitational action with counterterms in \cite{Emparan:1999pm}. The resulting energy,     $E_{\text{ren}} = M +  E_k^0   $, differs from the mass parameter by a constant term, the Casimir energy of the dual field theory 
 \begin{equation}
E_k^0 = \frac{\Omega_{k,d-1}L^{d-2}}{8\pi G} \epsilon_k^0 ,   
 \end{equation} 
 with  $\epsilon^0_k =0$ for odd $d$ and equal to \cite{Emparan:1999pm}
   \begin{equation} 
 	\epsilon_k^0 = (-k)^{d/2} \frac{(d-1)!!^2}{d!} \qquad \text{for even $d$}.
 \end{equation} 
For instance, $\epsilon_k^0=-k/2$ for $d=2$ and $\epsilon_k^0=3 k^2 /8$ for $d=4.$ The renormalized version of the   Smarr formula    reads
 \begin{equation}
  E_{\text{ren}}  =  	\frac{d-1}{d-2}  TS -  \frac{1}{d-2}\frac{  \Theta_{\text{ren}} \Lambda}{ 4\pi G},
 \end{equation}
 with a new (counterterm subtracted) Killing volume  
 \begin{equation}
 	\Theta_{\text{ren}} = - \frac{\Omega_{k,d-1}}{d} \left ( r_+^d  - \frac{d-2}{ d-1 }  L^d \epsilon_k^0 \right).
 \end{equation}
 The holographic Euler equation still takes the form
 \begin{equation}
 	E_{\text{ren}}  = T S + \mu_{\text{ren}} C,
 \end{equation}
 since the Casimir energy is also proportional to the central charge, which we normalize here as $C = \Omega_{k,d-1}L^{d-1}/16 \pi G$. 
But the chemical potential   is not given by Eq. \eqref{chemicalstatic} in the main text  anymore, since it receives a constant contribution from the vacuum energy
 \begin{equation} \label{renormchem}
 	\mu_{\text{ren}}     =  - \frac{r_+^{d-2}}{L^{d-1}} \left(\frac{r_+^2}{L^2}- k \right)+ \frac{2}{L} \epsilon_k^0 .
 \end{equation}
 For $d=2$ we find the chemical potential $\mu_{\text{ren}} = -   r_+^2 /L^3  $, which agrees with the expression found in Appendix \ref{app2dcft} (for $r_-=0$ and $R = L$).   
 For planar black holes ($k=0$) or very large hyperbolic or spherical black holes (with $r_+ \gg L$), the Casimir energy is effectively zero and hence there is no distinction between the renormalized energy and the vacuum-subtracted energy. As can be seen from \eqref{appenergy} and \eqref{renormchem}, there are additional thermodynamic relations  for these black holes  
   \begin{equation}
 	E   = -(d-1) \mu C \qquad \text{and} \qquad T S = - d\,  \mu C, 
 \end{equation}
 consistent with the infinite-volume limit   of Appendix \ref{appA}.

     \bibliography{chemistry}

\end{document}